\documentclass[preprintnumbers,amsmath,amssymbm,prd]{revtex4}
\usepackage{epsfig}
\usepackage{graphicx}
\usepackage{amssymb}

\begin{document}

\title{Analytic study of self-gravitating polytropic spheres with light rings}
\author{Shahar Hod}
\affiliation{The Ruppin Academic Center, Emeq Hefer 40250, Israel}
\affiliation{ } \affiliation{The Hadassah Institute, Jerusalem
91010, Israel}
\date{\today}

\begin{abstract}

\ \ \ Ultra-compact objects describe horizonless solutions of the
Einstein field equations which, like black-hole spacetimes, possess
null circular geodesics (closed light rings). We study {\it
analytically} the physical properties of spherically symmetric
ultra-compact isotropic fluid spheres with a polytropic equation of
state. It is shown that these spatially regular horizonless
spacetimes are generally characterized by two light rings
$\{r^{\text{inner}}_{\gamma},r^{\text{outer}}_{\gamma}\}$ with the
property ${\cal C}(r^{\text{inner}}_{\gamma})\leq{\cal
C}(r^{\text{outer}}_{\gamma})$, where ${\cal C}\equiv m(r)/r$ is the
dimensionless compactness parameter of the self-gravitating matter
configurations. In particular, we prove that, while black-hole
spacetimes are characterized by the lower bound ${\cal
C}(r^{\text{inner}}_{\gamma})\geq1/3$, horizonless ultra-compact
objects may be characterized by the opposite dimensionless relation
${\cal C}(r^{\text{inner}}_{\gamma})\leq1/4$. Our results provide a
simple analytical explanation for the interesting numerical results
that have recently presented by Novotn\'y, Hlad\'ik, and Stuchl\'ik
[Phys. Rev. D {\bf 95}, 043009 (2017)].
\end{abstract}
\bigskip
\maketitle

\section{Introduction}

Curved spacetimes describing highly compact astrophysical objects
may be characterized, according to the Einstein field equations, by
null circular geodesics (closed light rings) \cite{Bar,Chan,ShTe} on
which photons and gravitons can orbit the central self-gravitating
compact object. These null orbits are interesting from both a
theoretical and an astrophysical points of view and their physical
properties have been studied extensively by physicists and
mathematicians during the last five decades (see
\cite{Bar,Chan,ShTe,Pod,Ame,Ste,Goe,Mas,Dol,Dec,CarC,Hodf,Hodt1,Hodt2}
and references therein).

As demonstrated in \cite{Pod,Ame}, the optical appearance of a
highly compact collapsing star is determined by the physical
properties of its null circular geodesic \cite{Pod,Ame}. Likewise,
the intriguing phenomenon of strong gravitational lensing by highly
compact objects is related to the presence of light rings in the
corresponding curved spacetimes \cite{Ste}. In addition, as
explicitly shown in \cite{Goe,Mas,Dol,Dec,CarC,Hodf}, the discrete
quasinormal resonant spectra of compact astrophysical objects are
related, in the eikonal limit, to the physical properties (the
circulation time and the characteristic instability time scale) of
the null circular geodesics that characterize the corresponding
curved spacetimes \cite{Notekonon,Konon,Zde2n}.

Interestingly, it has recently been proved that spherically
symmetric black-hole spacetimes must posses at least one light ring
\cite{Hodlb}. In particular, the theorem presented in \cite{Hodlb}
has revealed the fact that the innermost null circular geodesic of
an asymptotically flat black hole must be located in a highly
compact spacetime region which is characterized by the dimensionless
lower bound \cite{Hodlb,Noteunits}
\begin{equation}\label{Eq1}
{{m(r^{\text{in}}_{\gamma})}\over{r^{\text{in}}_{\gamma}}}\geq
{1\over3}\ \ \ \ \text{for black holes}\  ,
\end{equation}
where $r^{\text{in}}_{\gamma}$ and $m(r^{\text{in}}_{\gamma})$ are
respectively the innermost (smallest) radius of the light ring which
characterizes the black-hole spacetime and the total gravitational
mass contained within this sphere.

Ultra-compact objects, spatially regular {\it horizonless} matter
configurations which, like black-hole spacetimes, possess light
rings, have attracted much attention in recent years as possible
exotic alternatives to the canonical black-hole spacetimes
\cite{Kei,Hodt0,Mag,CBH,Carm,Hodmz1,Hodmz2,Zde1}. In particular, in
a very interesting work, Novotn\'y, Hlad\'ik, and Stuchl\'ik
\cite{Zde1} (see also \cite{Zde2n}) have recently studied
numerically the physical properties of spherically symmetric
self-gravitating isotropic fluid spheres with a polytropic
pressure-density equation of state of the form \cite{Hdt}
\begin{equation}\label{Eq2}
p(\rho)=k_{\text{p}}\rho^{1+1/n}\  ,
\end{equation}
where the dimensionless physical parameter $n$ is the polytropic
index of the fluid system \cite{Hdt}. It is worth emphasizing that
the self-gravitating ultra-compact trapping polytropic spheres were
first mentioned in \cite{Stuch2n}.

Interestingly, it has been explicitly demonstrated numerically in
\cite{Zde1} that spatially regular polytropic spheres may possess
{\it two} light rings
$\{r^{\text{in}}_{\gamma},r^{\text{out}}_{\gamma}\}$ (see
\cite{CBH,Hodmz2} for related discussions) which are characterized
by the compactness inequality
\begin{equation}\label{Eq3}
{{m(r^{\text{in}}_{\gamma})}\over{r^{\text{in}}_{\gamma}}}\leq
{{m(r^{\text{out}}_{\gamma})}\over{r^{\text{out}}_{\gamma}}}\  .
\end{equation}
In particular, the intriguing fact has been revealed in \cite{Zde1}
that the spherically symmetric self-gravitating horizonless
polytropic spheres may be characterized by closed light rings (null
circular geodesics) with the remarkably small compactness relation
\begin{equation}\label{Eq4}
{{m(r^{\text{in}}_{\gamma})}\over{r^{\text{in}}_{\gamma}}}<
{1\over3}\ \ \ \ \text{for horizonless polytropic spheres}\  .
\end{equation}
It is worth emphasizing the fact that the dimensionless relation
(\ref{Eq4}), observed numerically in \cite{Zde1} for the {\it
horizonless} self-gravitating polytropic spheres, violates the lower
bound (\ref{Eq1}) which, as explicitly proved in \cite{Hodlb},
characterizes the innermost light rings of spherically symmetric
black-hole spacetimes.

The main goal of the present paper is to study analytically the
physical and mathematical properties of the horizonless
ultra-compact polytropic matter configurations. In particular, below
we shall provide compact {\it analytical} proofs for the
characteristic intriguing relations (\ref{Eq3}) and (\ref{Eq4}) that
have recently been observed {\it numerically} in \cite{Zde1} for the
spherically symmetric spatially regular isotropic fluid stars.

\section{Description of the system}

Following the interesting physical model studied numerically in
\cite{Zde1}, we shall consider asymptotically flat isotropic matter
configurations which are characterized by the spherically symmetric
static line element \cite{Chan,Notesc}
\begin{equation}\label{Eq5}
ds^2=-e^{-2\delta}\mu dt^2 +\mu^{-1}dr^2+r^2(d\theta^2 +\sin^2\theta
d\phi^2)\  ,
\end{equation}
where $\delta=\delta(r)$ and $\mu=\mu(r)$. Spatially regular matter
configurations are characterized by the functional behavior
\cite{Hodt1}
\begin{equation}\label{Eq6}
\mu(r\to 0)=1+O(r^2)\ \ \ \ {\text{and}}\ \ \ \ \delta(0)<\infty\
\end{equation}
in the near-origin $r\to0$ limit. In addition, asymptotically flat
regular spacetimes are characterized by the simple large-$r$
functional relations \cite{Hodt1,May}
\begin{equation}\label{Eq7}
\mu(r\to\infty) \to 1\ \ \ \ {\text{and}}\ \ \ \ \delta(r\to\infty)
\to 0\ .
\end{equation}

The non-linearly coupled Einstein-matter field equations,
$G^{\mu}_{\nu}=8\pi T^{\mu}_{\nu}$, can be expressed by the
differential relations \cite{Hodt1,Noteprm}
\begin{equation}\label{Eq8}
\mu'=-8\pi r\rho+{{1-\mu}\over{r}}\
\end{equation}
and
\begin{equation}\label{Eq9}
\delta'=-{{4\pi r(\rho +p)}\over{\mu}}\  ,
\end{equation}
where the radially-dependent density and pressure functions
\begin{equation}\label{Eq10}
\rho\equiv -T^{t}_{t}\ \ \ \ \text{and}\ \ \ \ p\equiv
T^{r}_{r}=T^{\theta}_{\theta}=T^{\phi}_{\phi}
\end{equation}
denote the components of the isotropic energy-momentum tensor
\cite{Bond1}. We shall assume that the spherically symmetric
asymptotically flat matter configurations respect the dominant
energy condition \cite{HawEl}
\begin{equation}\label{Eq11}
0\leq |p|\leq\rho\  .
\end{equation}

From the Einstein equations (\ref{Eq8}) and (\ref{Eq9}) and the
conservation relation
\begin{equation}\label{Eq12}
T^{\mu}_{r ;\mu}=0\  ,
\end{equation}
one can derive the characteristic compact differential equation
\begin{equation}\label{Eq13}
P'(r)= {{r}\over{2\mu}}\big[{\cal R}(\rho+p)+2\mu(-\rho+p)\big]\
\end{equation}
for the gradient of the radially-dependent isotropic pressure
function
\begin{equation}\label{Eq14}
P(r)\equiv r^2p(r)\  ,
\end{equation}
where
\begin{equation}\label{Eq15}
{\cal R}(r)\equiv 3\mu-1-8\pi r^2p\  .
\end{equation}

Below we shall analyze the spatial behavior of the characteristic
dimensionless compactness function
\begin{equation}\label{Eq16}
{\cal C}(r)\equiv {{m(r)}\over{r}}\  ,
\end{equation}
where the mass $m(r)$ of the matter fields contained within a sphere
of radius $r$ is given by the simple integral relation \cite{Hodt1}
\begin{equation}\label{Eq17}
m(r)=4\pi\int_{0}^{r} x^{2} \rho(x)dx\  .
\end{equation}
Taking cognizance of Eqs. (\ref{Eq8}) and (\ref{Eq17}), one deduces
the simple dimensionless functional relation
\begin{equation}\label{Eq18}
\mu(r)=1-{{2m(r)}\over{r}}\  .
\end{equation}
For later purposes we note that asymptotically flat regular matter
configurations are characterized by the asymptotic radial behavior
\cite{Hodt1}
\begin{equation}\label{Eq19}
r^3p(r)\to 0\ \ \ \ \text{for}\ \ \ \ r\to\infty\  .
\end{equation}

\section{Null circular geodesics of spherically symmetric curved spacetimes}

In the present section we shall follow the analysis presented in
\cite{Chan,CarC,Hodt1} in order to determine the radii of the null
circular geodesics (closed light rings) which characterize the
spherically symmetric self-gravitating ultra-compact objects. We
first note that the energy $E$ and the angular momentum $L$ provide
two conserved physical parameters along the null geodesics of the
static spacetime (\ref{Eq5}) \cite{Chan,CarC,Hodt1}.

In particular, the effective radial potential
\cite{Chan,CarC,Hodt1,Notedot}
\begin{equation}\label{Eq20}
E^2-V_r\equiv \dot
r^2=\mu\Big({{E^2}\over{e^{-2\delta}\mu}}-{{L^2}\over{r^2}}\Big)\
\end{equation}
determines, through the relations \cite{Chan,CarC,Hodt1,Notethr}
\begin{equation}\label{Eq21}
V_r=E^2\ \ \ \ \text{and}\ \ \ \ V'_r=0\  ,
\end{equation}
the null circular trajectories (light rings) of the static spacetime
(\ref{Eq5}). Substituting Eqs. (\ref{Eq8}), (\ref{Eq9}), and
(\ref{Eq20}) into (\ref{Eq21}), one obtains the characteristic
functional relation \cite{Chan,CarC,Hodt1}
\begin{equation}\label{Eq22}
{\cal R}(r=r_{\gamma})=0\
\end{equation}
for the null circular geodesics of the
spherically symmetric ultra-compact objects.

\section{An analytical proof of the characteristic relation
${\cal C}(r^{\text{in}}_{\gamma})\leq{\cal
C}(r^{\text{out}}_{\gamma})$ for horizonless isotropic ultra-compact
objects}

The physical properties of spherically symmetric self-gravitating
isotropic ultra-compact objects have recently been studied
numerically in the interesting work of Novotn\'y, Hlad\'ik, and
Stuchl\'ik \cite{Zde1} (see also \cite{Zde2n}). Intriguingly, it has
been explicitly shown in \cite{Zde1} that the horizonless curved
spacetimes of these spatially regular compact matter configurations
generally possess two light rings
$\{r^{\text{in}}_{\gamma},r^{\text{out}}_{\gamma}\}$ (see
\cite{CBH,Hodmz2} for related studies) which are characterized by
the dimensionless compactness relation (\ref{Eq3}).

In the present section we shall use {\it analytical} techniques in
order to provide a compact proof for the intriguing property ${\cal
C}(r^{\text{in}}_{\gamma})<{\cal C}(r^{\text{out}}_{\gamma})$ [see
Eqs. (\ref{Eq3}) and (\ref{Eq16})] which characterizes the
horizonless isotropic ultra-compact objects. We first point out
that, taking cognizance of Eqs. (\ref{Eq6}), (\ref{Eq7}),
(\ref{Eq15}), and (\ref{Eq19}), one finds the simple asymptotic
relations
\begin{equation}\label{Eq23}
{\cal R}(r=0)=2\ \ \ \ \text{and}\ \ \ \ {\cal R}(r\to\infty)\to 2\
\end{equation}
for the dimensionless radial function ${\cal R}(r)$. From Eqs.
(\ref{Eq22}) and (\ref{Eq23}) one deduces that, for horizonless
ultra-compact matter configurations with non-degenerate light rings
\cite{CBH,Hodmz2,Noteep}, the function ${\cal R}(r)$ is
characterized by the inequality
\begin{equation}\label{Eq24}
{\cal R}(r)<0\ \ \ \ \text{for}\ \ \ \ r\in
(r^{\text{in}}_{\gamma},r^{\text{out}}_{\gamma})\
\end{equation}
in the radial region between the two light rings of the
ultra-compact objects.

Substituting the characteristic inequality (\ref{Eq24}) into Eq.
(\ref{Eq13}) and taking cognizance of the relation (\ref{Eq11}), one
finds that $P(r)$ is a monotonically decreasing function between the
two light rings of the horizonless compact object:
\begin{equation}\label{Eq25}
P'(r)<0\ \ \ \ \text{for}\ \ \ \ r\in
(r^{\text{in}}_{\gamma},r^{\text{out}}_{\gamma})\  .
\end{equation}
In particular, from Eqs. (\ref{Eq14}), (\ref{Eq15}), (\ref{Eq22}),
and (\ref{Eq25}), one deduces that the dimensionless function
$\mu(r)$ is characterized by the inequality
\begin{equation}\label{Eq26}
\mu(r^{\text{in}}_{\gamma})\geq\mu(r^{\text{out}}_{\gamma})\  ,
\end{equation}
or equivalently [see Eqs. (\ref{Eq16}) and (\ref{Eq18})]
\begin{equation}\label{Eq27}
{\cal C}(r^{\text{in}}_{\gamma})\leq{\cal
C}(r^{\text{out}}_{\gamma})\ .
\end{equation}
We have therefore provided a simple analytical proof for the
numerically observed relation (\ref{Eq3}) \cite{Zde1} which
characterizes the horizonless isotropic ultra-compact objects.

\section{Upper bound on the compactness of the inner light ring of isotropic
ultra-compact objects}

The characteristic compactness parameter ${\cal C}(r)\equiv m(r)/r$
of the self-gravitating ultra-compact objects can be computed using
the numerical procedure described in \cite{Zde1}. Intriguingly, as
demonstrated explicitly in \cite{Zde1}, the spatially regular
horizonless ultra-compact objects may be characterized by inner
light rings whose dimensionless compactness parameter ${\cal
C}(r^{\text{in}}_{\gamma})$ is well below the lower bound
(\ref{Eq1}) which, as explicitly proved in \cite{Hodlb},
characterizes the innermost null circular geodesics (light rings) of
spherically symmetric asymptotically flat black-hole spacetimes.

In Table \ref{Table1} we present, for various values of the
polytropic index $n$, the numerically computed dimensionless
compactness parameter ${\cal
C}^{\text{numerical}}(r^{\text{in}}_{\gamma})$ of the isotropic
ultra-compact objects \cite{Zde1,Notecmx,Too}. One finds that ${\cal
C}(r^{\text{in}}_{\gamma};n)$ is a monotonically decreasing function
of the dimensionless polytropic index $n$. Interestingly, we find
that the numerical results presented in Table \ref{Table1} are
described extremely well by the simple asymptotic formula (see Table
\ref{Table1})
\begin{equation}\label{Eq28}
{\cal C}(r^{\text{in}}_{\gamma};n)=\alpha
+{{\beta}\over{n}}+O(n^{-2}) \ \ \ \ \text{with}\ \ \ \
\alpha=0.2149 \ \ \ \text{and}\ \ \ \beta=0.1602\  .
\end{equation}

\begin{table}[htbp]
\centering
\begin{tabular}{|c|c|c|}
\hline $\ \text{Polytropic}\ $\ \ & \ ${\cal
C}^{\text{numerical}}(r^{\text{in}}_{\gamma})$ \ &
\ ${\cal C}^{\text{analytical}}(r^{\text{in}}_{\gamma})$ \ \\
$\ \ \text{index}\ \ n\ \ $\ \ & \ \ $\text{Ref.}$\ \cite{Zde1} \ \
&
\ \ $\text{Eq.}$\ (\ref{Eq28}) \ \ \\
\hline
\ \ $2.2$\ \ \ &\ \ $0.2906$\ \ &\ \ $0.2877$\ \ \\
\ \ $2.4$\ \ \ &\ \ $0.2824$\ \ &\ \ $0.2817$\ \ \\
\ \ $2.6$\ \ \ &\ \ $0.2767$\ \ &\ \ $0.2765$\ \ \\
\ \ $2.8$\ \ \ &\ \ $0.2723$\ \ &\ \ $0.2721$\ \ \\
\ \ $3.0$\ \ \ &\ \ $0.2683$\ \ &\ \ $0.2683$\ \ \\
\ \ $3.2$\ \ \ &\ \ $0.2649$\ \ &\ \ $0.2650$\ \ \\
\ \ $3.4$\ \ \ &\ \ $0.2620$\ \ &\ \ $0.2620$\ \ \\
\ \ $3.6$\ \ \ &\ \ $0.2594$\ \ &\ \ $0.2594$\ \ \\
\ \ $3.8$\ \ \ &\ \ $0.2570$\ \ &\ \ $0.2571$\ \ \\
\ \ $4.0$\ \ \ &\ \ $0.2549$\ \ &\ \ $0.2550$\ \ \\
\hline
\end{tabular}
\caption{Ultra-compact polytropic fluid spheres with light rings. We
present, for various values of the polytropic index $n$, the {\it
numerically} computed \cite{Zde1,Notecmx,Too} dimensionless
compactness parameter ${\cal
C}^{\text{numerical}}(r^{\text{in}}_{\gamma};n)$ of the isotropic
matter configurations. We also present the corresponding values of
the dimensionless compactness parameter ${\cal
C}^{\text{analytical}}(r^{\text{in}}_{\gamma};n)$ as calculated
directly from the simple analytical fit (\ref{Eq28}). One finds a
remarkably good agreement between the numerical results \cite{Zde1}
and the analytical formula (\ref{Eq28}). In particular, one deduces
from (\ref{Eq28}) the characteristic asymptotic value ${\cal
C}(r^{\text{in}}_{\gamma})\to 0.2149<1/4$ for $n\gg1$.}
\label{Table1}
\end{table}

What we find most interesting is the fact that the horizonless
ultra-compact isotropic objects are characterized by the
dimensionless asymptotic compactness parameter
\begin{equation}\label{Eq29}
{\cal C}(r^{\text{in}}_{\gamma};n\gg1)\simeq0.2149<1/4\  .
\end{equation}
As emphasized above, one immediately realizes that the asymptotic
value (\ref{Eq29}), which characterizes the spatially regular
horizonless matter configurations, is well below the lower bound
${\cal C}(r_{\gamma})\geq 1/3$ [see Eqs. (\ref{Eq1}) and
(\ref{Eq16})] which characterizes the corresponding spherically
symmetric black-hole spacetimes \cite{Hodlb}.

In the present section we shall provide an {\it analytical}
explanation for the {\it numerically} inferred asymptotic behavior
(\ref{Eq29}) of the dimensionless compactness parameter. In
particular, we shall now derive an upper bound on the compactness
parameter ${\cal C}(r^{\text{in}}_{\gamma};n)$ of the isotropic
ultra-compact objects in the $n\gg1$ \cite{Notestuch2n,Stuch2n}
limit of the polytropic index, which corresponds to the limiting
pressure-density relation [see Eq. (\ref{Eq2})]
\begin{equation}\label{Eq30}
p=k_{\text{p}}\rho\  .
\end{equation}

From Eqs. (\ref{Eq8}), (\ref{Eq13}), (\ref{Eq14}), and (\ref{Eq15}),
one finds the gradient relation
\begin{equation}\label{Eq31}
{\cal R}'(r=r_{\gamma})={{2}\over {r_{\gamma}}}\big[1-8\pi
r^2_{\gamma}(\rho+p)\big]\
\end{equation}
for the isotropic ultra-compact objects. In addition, taking
cognizance of Eqs. (\ref{Eq22}) and (\ref{Eq23}), one deduces that
the inner light ring of a spatially regular ultra-compact
horizonless matter configuration is characterized by the relation
${\cal R}'(r=r^{\text{in}}_{\gamma})\leq 0$ \cite{Notegnh}, or
equivalently [see Eq. (\ref{Eq31})]
\begin{equation}\label{Eq32}
8\pi r^2_{\gamma}(\rho+p)\geq1\ \ \ \ \text{for}\ \ \ \
r=r^{\text{in}}_{\gamma}\  .
\end{equation}

Taking cognizance of Eqs. (\ref{Eq15}), (\ref{Eq30}), and
(\ref{Eq32}), one obtains the lower bound
\begin{equation}\label{Eq33}
\mu(r^{\text{in}}_{\gamma};n\gg1)\geq
{{2k_{\text{p}}+1}\over{3(k_{\text{p}}+1)}}\ ,
\end{equation}
which yields the characteristic upper bound [see Eqs. (\ref{Eq16})
and (\ref{Eq18})]
\begin{equation}\label{Eq34}
{\cal C}(r^{\text{in}}_{\gamma};n\gg1)\leq
{{k_{\text{p}}+2}\over{6(k_{\text{p}}+1)}}\
\end{equation}
on the dimensionless compactness parameter which characterizes the
inner null circular geodesic (inner light ring) of the isotropic
ultra-compact objects. In particular, taking cognizance of Eqs.
(\ref{Eq11}) and (\ref{Eq30}), one deduces that, in the $n\gg1$
limit of the polytropic index, the physical parameter $k_{\text{p}}$
is bounded from above by the simple relation $k_{\text{p}}\leq1$.
Substituting the limiting value $k_{\text{p}}\to1^-$ into
(\ref{Eq34}), one obtains the characteristic upper bound
\cite{Notekp0,Notegrc}
\begin{equation}\label{Eq35}
{\cal C}(r^{\text{in}}_{\gamma};n\gg1,k_{\text{p}}\to1^-)<
{1\over4}\ .
\end{equation}

It is worth emphasizing the fact that the {\it analytically} derived
upper bound (\ref{Eq35}) on the dimensionless compactness parameter
is consistent with the asymptotic behavior (\ref{Eq29}) which stems
from the {\it numerical} studies \cite{Zde1} of the self-gravitating
ultra-compact isotropic fluid configurations. In particular, in this
section we have provided an explicit {\it analytical} proof to the
{\it numerically} observed intriguing fact that horizonless
ultra-compact objects can violate the lower bound (\ref{Eq1}) which
characterizes spherically symmetric black-hole spacetimes
\cite{Notebhs}.

\section{Summary}

Horizonless spacetimes describing self-gravitating ultra-compact
matter configurations with closed light rings (null circular
geodesics) have recently attracted much attention as possible
spatially regular exotic alternatives to canonical black-hole
spacetimes (see \cite{Kei,Hodt0,Mag,CBH,Carm,Hodmz1,Hodmz2,Zde1} and
references therein).

In particular, the physical properties of horizonless ultra-compact
isotropic fluid spheres with a polytropic equation of state have
recently been studied numerically in the physically important work
of Novotn\'y, Hlad\'ik, and Stuchl\'ik \cite{Zde1}. Interestingly,
it has been explicitly shown numerically in \cite{Zde1} that these
spherically symmetric spatially regular ultra-compact polytropic
matter configurations generally posses {\it two} closed light rings
(see also \cite{CBH,Hodmz2} for related discussions).

In the present paper we have used {\it analytical} techniques in
order to explore the physical and mathematical properties of the
ultra-compact polytropic stars. In particular, it has been
explicitly proved that the two light rings of these horizonless
matter configurations are characterized by the relation [see Eqs.
(\ref{Eq16}) and (\ref{Eq27})]
\begin{equation}\label{Eq36}
{\cal C}(r^{\text{in}}_{\gamma})\leq{\cal
C}(r^{\text{out}}_{\gamma})\ .
\end{equation}

Interestingly, we have further proved that, while spherically
symmetric black-hole spacetimes are characterized by the lower bound
${\cal C}(r^{\text{in}}_{\gamma})\geq1/3$ [see Eqs. (\ref{Eq1}) and
(\ref{Eq16})] \cite{Hodlb}, the spatially regular horizonless
ultra-compact objects are characterized by the opposite
dimensionless relation
\begin{equation}\label{Eq37}
{\cal C}(r^{\text{in}}_{\gamma};n\gg1,k_{\text{p}}\to1^-)<
{1\over4}\ .
\end{equation}
It is worth noting that the analytically derived upper bound
(\ref{Eq37}) on the characteristic dimensionless compactness
parameter ${\cal C}(r^{\text{in}}_{\gamma})$ is consistent with the
numerically inferred asymptotic behavior (\ref{Eq29}).

Finally, it is interesting to emphasize the fact that the analytical
results derived in the present paper provide a simple {\it
analytical} explanation for the interesting {\it numerical} results
that have recently presented by Novotn\'y, Hlad\'ik, and Stuchl\'ik
\cite{Zde1} for the physical properties of the self-gravitating
ultra-compact polytropic spheres.

\bigskip
\noindent {\bf ACKNOWLEDGMENTS}

This research is supported by the Carmel Science Foundation. I would
like to thank Yael Oren, Arbel M. Ongo, Ayelet B. Lata, and Alona B.
Tea for stimulating discussions.


\begin{thebibliography}{99}
\bibitem{Notekonon} It is worth mentioning that, as shown in \cite{Konon},
there are some non-trivial situations in which this relation is
violated. However, as shown in \cite{Zde2n}, this relation holds
true for the trapping polytropic spheres that we shall consider in
the present paper.
\bibitem{Noteunits} We shall use natural units in which $G=c=1$.
\bibitem{Notesc} Here $(t,r,\theta,\phi)$ are the familiar Schwarzschild spacetime coordinates.
\bibitem{Notethr} Note that these two functional relations, which determine the null circular
geodesics of the spherically symmetric spacetime (\ref{Eq5}),
correspond to $\dot r^2=(\dot r^2)'=0$ \cite{Chan,CarC,Hodt1}.
\bibitem{Noteep} It is important to note that, as explicitly
shown in \cite{Hodmz2}, there are special horizonless matter
configurations with degenerate light rings, which are characterized
by the relations ${\cal R}(r=r_{\gamma})={\cal R}'(r=r_{\gamma})=0$,
that may violate the inequality (\ref{Eq24}). We shall henceforth
consider in this section generic \cite{CBH,Hodmz2} self-gravitating
ultra-compact objects which respect the relation (\ref{Eq24}).
\bibitem{Notecmx} The numerical results presented in Table
\ref{Table1} correspond to self-gravitating isotropic fluid spheres
which are characterized by the limiting pressure-to-density ratio
$p_{\text{c}}/\rho_{\text{c}}=n/(n+1)$ at their centers \cite{Zde1}.
These matter configurations saturate the upper bound
$v_{\text{s}}\leq1$ on the speed of sound which is imposed by
causality requirements \cite{Too}.
\bibitem{Notegnh} It is important to emphasize that in the present section
we consider generic spatially regular field configurations with
light rings [that is, we consider the case of matter configurations
with non-degenerate light rings as well as the special case of
matter configurations with degenerate light rings for which ${\cal
R}'(r=r_{\gamma})=0$].
\bibitem{Notekp0} It is worth noting that the upper bound (\ref{Eq33})
on the physically allowed values of the dimensionless compactness
parameter ${\cal C}(r^{\text{in}}_{\gamma};n\gg1)$ is a
monotonically decreasing function of the physical parameter
$k_{\text{p}}$. For example, substituting the opposite limit
$k_{\text{p}}\to0^+$ into (\ref{Eq33}), one obtains the upper bound
${\cal C}(r^{\text{in}}_{\gamma};n\gg1,k_{\text{p}}\to0^+)<1/3$. The
numerical results presented in Table \ref{Table1} correspond to
horizonless ultra-compact isotropic fluid spheres which are
characterized by the limiting central ratio
$p_{\text{c}}/\rho_{\text{c}}=n/(n+1)$ which is imposed by causality
requirements \cite{Zde1,Too}. In particular, in the $n\gg1$ regime
one finds the limiting behavior $p_{\text{c}}/\rho_{\text{c}}\to1^-$
for the physically acceptable matter configurations, which
corresponds to the limiting value $k_{\text{p}}\to1^-$ [see Eq.
(\ref{Eq30})].
\bibitem{Notegrc} It is important to emphasize that, in the analytical
derivation of the upper bound (\ref{Eq35}), we have {\it not} used
the inequality (\ref{Eq24}) which, as explicitly demonstrated in
\cite{Hodmz2}, may be violated by special ultra-compact matter
configurations with degenerate light rings \cite{Noteep}.
\end{thebibliography}
\end{document}